\documentclass[12pt]{article}

\usepackage[dvips]{graphics}
\usepackage{amsmath, amsthm, amssymb}
\usepackage{graphicx}
\usepackage{times}
\usepackage{dcolumn}
\usepackage{bm}

\newcommand{\bra}{\langle}

\newcommand{\ket}{\rangle}

\usepackage{graphicx}
\usepackage{qcmc06}

\begin{document}

\pagestyle{empty}

\title{Quantum phase transitions with Photons and Polaritons}

\author{Fernando G.S.L. Brand\~ao, Michael J. Hartmann, 
        and Martin B. Plenio}

\affiliation{Institute for Mathematical Sciences, Prince's Gate, 53 Exhibition Road, Imperial College London, London SW7 2PG, United Kingdom\\
Quantum Optics and Laser Science group, Blackett Laboratory,
Imperial College London, London SW7 2BW}

\begin{abstract}

We show that a system of polaritons - combined atom and photon excitations - in an array of coupled cavities, under an experimental set-up usually considered in electromagnetically induced transparency, is described by the Bose-Hubbard model. This opens up the possibility of using this system as a quantum simulator, allowing for the observation of quantum phase transitions and for the measurement of local properties, such as single site observables. All the basic building blocks of the proposed setting have already been achieved experimentally, showing the feasibility of its realization in the near future.   

\end{abstract}

\setcounter{section}{0}

\section{Introduction}

The exponential growth of the Hilbert space dimension with the number of particles of the system is a hurdle for the classical simulation of quantum many-body systems, in particular near or at criticality. A possible solution to this conundrum, pioneered by Feynman 20 years ago \cite{Feynman}, is the use of quantum simulators, i.e. artificial quantum systems which generate the evolution of rather complex Hamiltonians in a controlled and systematic manner. An emblematic example of a quantum simulator is a quantum computer, capable of simulating any quantum evolution. Yet, the overwhelmingly high degree of accuracy required to build a useful quantum computer - operating below the fault-tolerance threshold - renders unlikely the availability of such a device in the near future. A natural approach is then to consider less general quantum simulators, engineered to simulate specific many-body models, but requiring much less stringent experimental conditions. 

Cold atoms trapped in an optical lattice constitutes an important example of this second kind of simulator \cite{Jaksch, Greiner, TonksG, Lewenstein}. By adjusting the experimental parameters properly, the Bose-Hubbard (BH) Hamiltonian - extensively used in modeling strongly correlated many-particle systems - can be generated. This idea, in turn, culminated in several important experiments emulating condensed-matter phenomena, such as the superfluid-to-Mott-insulator quantum phase transition \cite{Greiner} and the generation of a Tonks-Girardeau gas for the first time \cite{TonksG}. Moreover, the realization of the BH model opened up the possibility of studying several interesting aspects of many-body quantum systems, such as disorder, Anderson localization, and quantum magnetism \cite{Lewenstein}. Despite their impressive success, optical lattices are in all their applications limited by an intrinsic draw back: due to the separation between neighboring lattice sites of only half of the employed optical wavelength, it is extremely challenging to access them individually. 

Recently, we have proposed an alternative way to create an effective Bose-Hubbard model, which, in contrast to optical lattices, allows for the manipulation and measurement of the properties of individual constituent particles \cite{Hartmann} (see Refs. \cite{Angelakis} for related proposals). This proposal can be
realized in coupled arrays of optical cavities each interacting with an ensemble of atoms leading to the dynamical evolution of polaritons, combined atom-photon excitations, in this setting. An exciting feature of this proposal is the possibility of observing strong-correlation phenomena in a system of photons, which emerges as one of the limits of the polariton excitations.

\begin{figure}
\begin{center}
\includegraphics[scale=0.65]{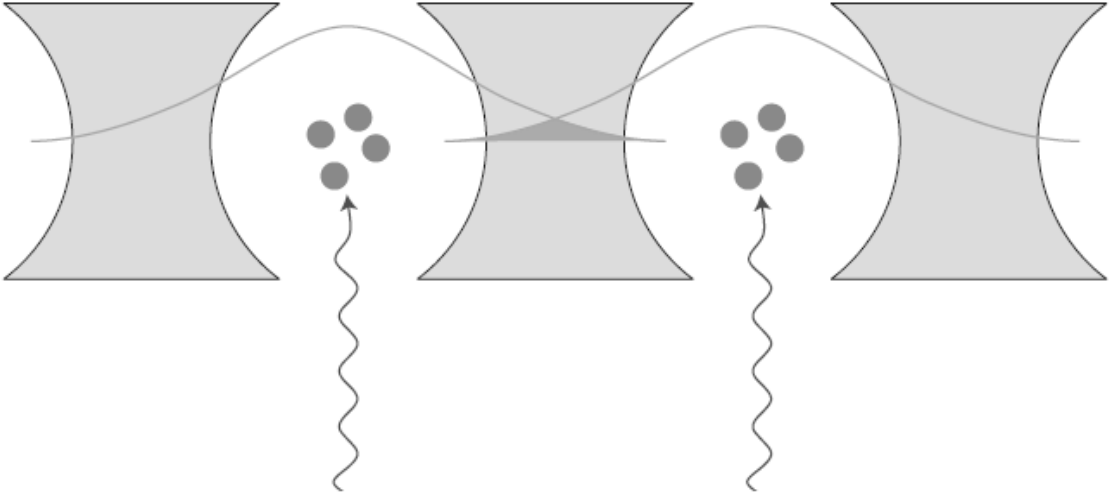}
\caption{\label{cavarr}(taken from Ref. \cite{Hartmann}) A schematic view of the model considered.}
\end{center}
\end{figure}

Our model consists of an array of cavities in an arbitrary geometry, where each cavity interacts with an ensemble of atoms, which are driven by an external laser. Photon hopping occurs between neighboring cavities due to the overlapp of their photonic wavefunctions (see fig. \ref{cavarr}) while
the repulsive force between two polaritons occupying the same cavity is generated by a large Kerr nonlinearity that occurs if atoms with
a specific level structure, usually considered in the context of Electromagnetically Induced Transparency
\cite{HFI90,ISWD97, WI99, Hau99,FIM05}, interact with light. By varying the intensity of the driving laser and hence the strength of the generated Kerr nonlinearity, the system can be driven through the superfluid-to-Mott-insulator transition. In particular, the driving laser may be adjusted for each cavity individually allowing for a much wider
range of tuning possibilities than in an optical lattice, including attractive interactions \cite{Hartmann}.

Our system can be described in terms of three species of
polaritons, where under the conditions specified below, the species which is least vulnerable
to decay processes is governed by the effective Bose-Hubbard (BH) Hamiltonian,
\begin{equation} \label{bosehubbard}
H_{\text{eff}} = \kappa \, \sum_{\vec{R}} \, \left(p_{\vec{R}}^{\dagger}\right)^2 \left(p_{\vec{R}}\right)^2 \, + \,
J \sum_{\bra \vec{R}, \vec{R}' \ket} \left(
p_{\vec{R}}^{\dagger} \, p_{\vec{R}'}\, + \, \text{h.c.} \right) \, .
\end{equation}
$p_{\vec{R}}^{\dagger}$ creates a polariton in the cavity at site $\vec{R}$ and the parameters $\kappa$ and $J$ describe on site repulsion and inter cavity hopping respectively.

\section{Derivation of the Model}

We start by considering a periodic array of cavities, which we describe here by a real, periodic 
dielectric constant.
The electromagnetic field can be expanded in Wannier functions,
$\vec{w}_{\vec{R}}$, each localized at one single cavity at location
$\vec{R}$. 
In terms of the creation and annihilation operators of the Wannier modes, $a_{\vec{R}}^{\dagger}$ and $a_{\vec{R}}$,
the Hamiltonian of the field can be written,
\begin{equation} \label{arrayham2}
\mathcal{H} = \omega_C \sum_{\vec{R}}
\left(  a_{\vec{R}}^{\dagger} a_{\vec{R}} + \frac{1}{2} \right) +
2 \omega_C \alpha \sum_{\bra \vec{R}, \vec{R}' \ket}
\left( a_{\vec{R}}^{\dagger} a_{\vec{R}'} + \text{h.c.} \right) \, .
\end{equation}
Here $\sum_{\bra \vec{R}, \vec{R}' \ket}$ is the sum of all pairs of cavities which are nearest neighbors
of each other.
$\alpha$ is given by  an overlap integral of neighboring Wannier functions \cite{YXLS99} and can be obtained numerically for specific models.
Since $\alpha \ll 1$, we neglected rotating terms which contain products of two creation or two annihilation operators of Wannier modes in deriving (\ref{arrayham2}).  The model (\ref{arrayham2}) provides an excellent approximation to many relevant implementations such as coupled photonic crystal micro-cavities or fiber coupled toroidal micro-cavities. Furthermore, it allows for the observation of state transfer and entanglement dynamics and propagation for Gaussian states \cite{HRP06, PHE04}.

The on site interaction potential for polaritons is generated in each cavity by atoms of a particular level structure that are driven with an external laser in the same manner as in Electromagnetically Induced Transparency,
see figure 2:
the transitions between levels 2-4 and 1-3 couple via dipole moments to the cavity resonance mode, whereas the transitions between levels 2 and 3 are coupled to the laser field.

\begin{figure}
\begin{center}
\includegraphics[scale=0.45]{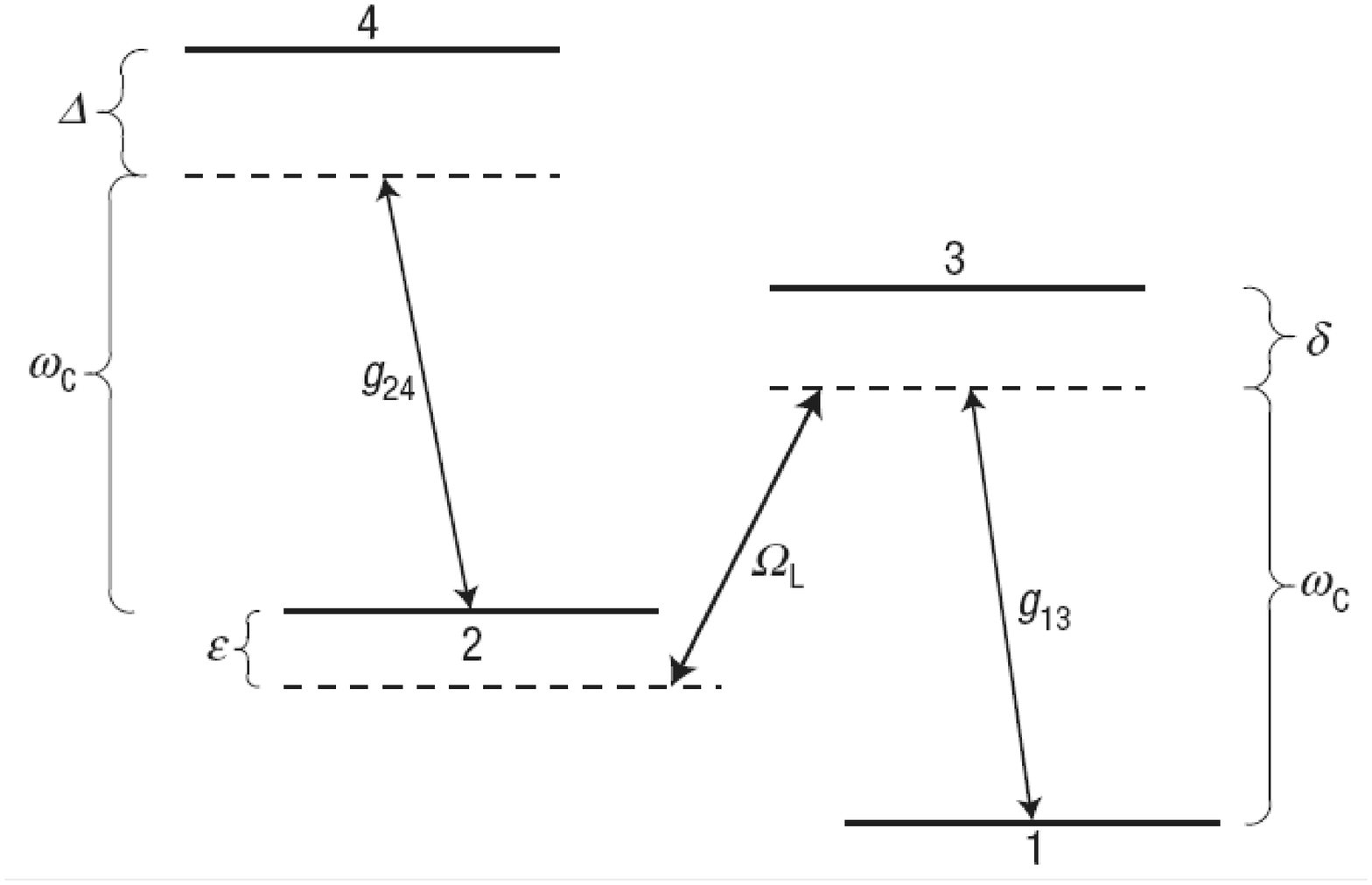}
\caption{(taken from Ref. \cite{Hartmann}) Diagram of the level structure considered.}
\end{center}
\end{figure}

In a rotating frame with respect to
$H_0 = \omega_C \left( a^{\dagger} a + \frac{1}{2} \right) + \sum_{j=1}^N \left( \omega_C \sigma_{22}^j + \omega_C \sigma_{33}^j + 2 \omega_C \sigma_{44}^j \right)$,
the Hamiltonian of the atoms in one cavity reads,
$H_I =
\sum_{j=1}^N \left(\varepsilon \sigma_{22}^j + \delta \sigma_{33}^j + (\Delta + \varepsilon)
\sigma_{44}^j \right) \, + \, \sum_{j=1}^N \left( \Omega_L \, \sigma_{23}^j \, + \,
g_{13} \, \sigma_{13}^j \, a^{\dagger} \, + \,
g_{24} \, \sigma_{24}^j \, a^{\dagger} \, + \, \text{h.c.} \right)$.
Here $\sigma_{kl}^j = | k_j \ket \bra l_j |$ projects level $l$ of atom $j$ to level $k$ of the same atom,
$a^{\dagger}$ creates one photon in the cavity,
$\omega_C$ is the frequency of the cavity mode,
$\Omega_L$ the Rabi frequency of the driving by the laser, $g_{13}$ and $g_{24}$
the parameters of the dipole coupling of the cavity mode to the respective atomic
transitions and $\Delta$, $\delta$ and $\varepsilon$ are detunings. All these parameters are assumed to be real.
 
Neglecting the coupling to level 4, $g_{24} = 0$, and two photon detuning, $\varepsilon = 0$, the Hamiltonian $H_I$ can
be written as a model of three species of polaritons \cite{WI99,FL00} with the creation (and annihilation) operators
$p_0^{\dagger}  = \frac{1}{B} \, \left(g S_{12}^{\dagger} - \Omega_L a^{\dagger} \right)$ and
$p_{\pm}^{\dagger} = \sqrt{\frac{2}{A (A \pm \delta)}} \, \left(\Omega_L S_{12}^{\dagger} + g a^{\dagger} \pm
\frac{A \pm \delta}{2} S_{13}^{\dagger} \right)$, 
where $g = \sqrt{N} g_{13}$, $B = \sqrt{g^2 + \Omega_L^2}$, $A = \sqrt{4 B^2 + \delta^2}$,
$S_{12}^{\dagger} = \frac{1}{\sqrt{N}} \sum_{j=1}^N \sigma_{21}^j$
and $S_{13}^{\dagger} = \frac{1}{\sqrt{N}} \sum_{j=1}^N \sigma_{31}^j$.
In the limit of large atom numbers, $N \gg 1$, $p_0^{\dagger}$, $p_+^{\dagger}$ and $p_-^{\dagger}$ satisfy bosonic commutation relations and for $g_{24} = 0$ and $\varepsilon = 0$ the Hamiltonian $H_I$ thus describes independent non-interacting bosonic particles,
\begin{equation} \label{H0polariton}
\left[H_I\right]_{g_{24} = 0, \varepsilon = 0} = 
\mu_0 \, p_0^{\dagger} p_0 + \mu_+ \, p_+^{\dagger} p_+ + \mu_- \, p_-^{\dagger} p_- \, ,
\end{equation}
where the polariton frequencies are given by $\mu_0 = 0$, $\mu_+ = (\delta - A)/2$ and
$\mu_- = (\delta + A)/2$. We will now focus on the dark state polaritons, $p_0^{\dagger}$,
which will be shown to be described by the Hamiltonian (\ref{bosehubbard}).

For $|g_{24}| \, , \, |\varepsilon| \, , \, |\Delta| \, \ll \, |\mu_+ - \mu_0| \, , \, |\mu_- - \mu_0|$,
the coupling $g_{24}$ and the two photon detuning $\varepsilon$ do not induce interactions between different polariton species \cite{Hartmann}.
Assuming further that $\sqrt{n_p (n_p - 1)} g_{24} \ll |\Delta|$, where $n_p$ is the number of polaritons $p_0^{\dagger}$, the coupling to level 4 can be treated perturbatively. The result is an energy shift of $n_p (n_p - 1)\kappa$ with 
\begin{equation}
\kappa = - \frac{g_{24}^2}{\Delta} \,
\frac{N g_{13}^2 \, \Omega_L^2}{\left(N g_{13}^2 \, + \, \Omega_L^2 \right)^2}.
\end{equation}
Note that $\kappa > 0$ for $\Delta < 0$ and vice versa.
The two photon detuning $\varepsilon$, in turn, leads analogously to an energy shift of
$\varepsilon \, g^2 \, B^{-2}$ for the polariton $p_0^{\dagger}$, which plays the role of a chemical
potential in the effective Hamiltonian.

The interaction between cavities, derived above in terms of the photonic operators, can be reexpressed in terms of the polaritons scpecies. Due to the large separation in the polaritons frequencies,
$|2 \omega_C \alpha| \, \ll \, |\mu_+ - \mu_0| \, , \, |\mu_- - \mu_0|$, the hopping does not mix different polariton species. The Hamiltonian for the polaritons $p_{\vec{R}}^{\dagger}$ (the dark state polariton $p_{0}^{\dagger}$ at site $\vec{R}$) thus takes on the form
(\ref{bosehubbard}), with 
\begin{equation}
J = \frac{2 \omega_C \Omega_L^2}{N g_{13}^2 + \Omega_L^2} \alpha \, ,
\end{equation}
where we have assumed a negligible two photon detuning, $\varepsilon \approx 0$.

The number of polaritons in one individual cavity can be measured via resonance fluorescence.
To that end, the polaritons are made purely atomic excitations by adiabatically switching off the driving
laser \cite{FL00}. The process is adiabatic if $g B^{-2} \frac{d}{dt} \Omega_L \ll |\mu_+| , |\mu_-|$ \cite{FIM05}, which means it can be done fast enough to prevent polaritons from hopping between
cavities during the switching. Hence, in each cavity, the final number of atomic excitations in level 2 is equal to the initial number of dark state polaritons. By measuring the number of polaritons in several individual cavities in possibly repeated resonance flourescence experiments one
can obtain the local number fluctuations which allow to distinguish between the different phases of the system.

\section{Feasibility of the observation of phase transitions}

Promising candidates for an experimental realisation are photonic band gap cavities \cite{AAS+03,SNAA05} and toroidal or spherical micro-cavities, which are coupled via tapered optical fibres \cite{AKS+03}. These cavities can be produced and positioned with high precision and in large numbers. They have a very large Q-factor ($> 10^8$) for light that is trapped as whispering gallery modes and efficient coupling to optical fibres \cite{YAV03} as well as coupling to Cs-atoms in close proximity to the cavity via the evanescent field \cite{ADW+06,BPK06} have been demonstrated experimentally. Photonic crystals represent an appealing alternative as they offer the possibility for the fabrication of large arrays of cavities in lattices or networks \cite{YXLS99,BHA+05}. These cavities have been realised with Q-factors of $10^6$ and higher Q-factors of $2 \times 10^7$ have been predicted \cite{SNAA05}.

As levels 1 and 2 are metastable, spontaneous emission can only occur from levels 3 and 4. However, due to pratically negligible population of these two levels, the main loss channel is cavity decay. For successfully observing the dynamics and phases of the effective Hamiltonian
(1), the repulsion term $\kappa$
needs to be much larger than the damping rate for the dark state polaritons, $\Gamma$ \cite{Hartmann}. Choosing $\Omega_L \ll \sqrt{N} g_{13}$, one can currently achieve a ratio of
$\kappa / \Gamma \sim 5.2$ for photonic band gap cavities
(for predicted Q-factors this can go up to $\kappa / \Gamma \sim 170$) \cite{SNAA05},
while for toroidal micro-cavities the impressive large value of $\kappa / \Gamma \sim 1.1 \times 10^{3}$
is possible \cite{SKV+05,ADW+06}, making both ideal candidates for an experimental implementation.

As an ilustration, a numerical simulation of the Mott-insulator-to-superfluid phase transition is shown in figure 3.
We consider three coupled cavities with periodic boundary conditions,
which are prepared such that there is initially one polariton $p_{\vec{R}}^{\dagger}$ in each. We take parameters for toroidal microcavities from \cite{SKV+05}, $g_{24} = g_{13} = 2.5 \times 10^{9} s^{-1}$, $\Gamma_3 = \Gamma_4 = 1.6 \times 10^{7} s^{-1}$ (spontaneous emittion rate from levels 3 and 4), $\Gamma_C = 0.4 \times 10^{5} s^{-1}$ (cavity decay rate), $N = 1000$, $\Delta = -2.0 \times 10^{10} s^{-1}$, and $2 \omega_c \alpha = 1.1 \times 10^{7} s^{-1}$. 
To drive the system through the transition, The Rabi frequency of the driving laser is imcreased from $\Omega_L = 7.8 \times 10^{10} s^{-1}$ to $\Omega_L = 1.1 \times 10^{12} s^{-1}$ throughout
the experiment, while all other parameters remain constant. Fig. 3 shows in the left part, $\Omega_L$ (inset), the on-site interaction strength $\kappa$ and the hopping rate $J$ during the transition.
The right part shows that number fluctuations $F_1 = \langle (p_1^{\dagger} p_1)^2 \rangle - \langle p_1^{\dagger} p_1 \rangle^2$ are small in the Mott phase and increase when driving into the superfluid phase. The average number of polaritons in one cavity $n_1 = \langle p_1^{\dagger} p_1 \rangle$ stays close to unity across the whole simulated time range confirming that polariton loss is indeed negligible on this time scale.

\begin{figure}
\begin{center}
\includegraphics[scale=0.5]{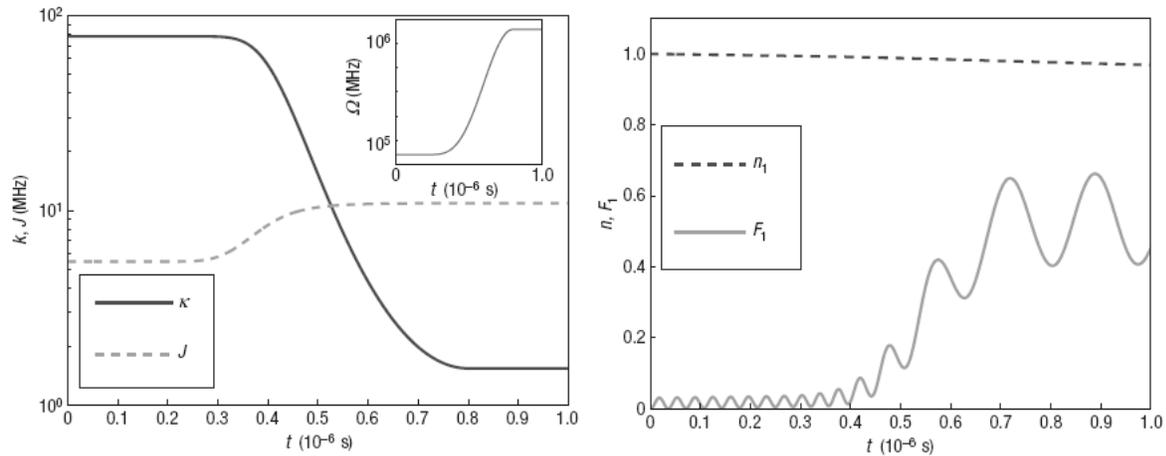}
\caption{(taken from Ref. \cite{Hartmann}) \textit{left}: Log plot of of $\kappa$ and $J$ and linear plot of the time-dependent $\Omega_L$ (inset). \textit{right}: expectation value ($n_1$) and fluctuations ($F_1$) for the number of polaritons in one site considering damping (see main text for details).}
\end{center}
\end{figure}

\section{Conclusion}

In summary, we have demonstrated, that polaritons, combined atom - photon excitations, can form an effective quantum many particle system, which is described by a Bose-Hubbard Hamiltonian. We have demonstrated numerically that the observation of the Mott insulator to superfluid phase transition is indeed feasible for experimentally realizable parameters. In contrast to earlier realizations, our scenario has the advantage that single sites of the BH model can be addressed individually and offers the possibility of creating a BH type Hamiltonian with an attractive on site potential.

\end{document}